\documentstyle[prb,aps,epsf]{revtex}

\draft
\title{Phonon mediated hole pairing in the 1D Hubbard model near half-filling}
\author{Yury Petrov} 
\address{Department of Physics and Astronomy, University of Pennsylvania, Pennsylvania 19104}
\author{Takeshi Egami}
\address{Department of Materials Science and Engineering, University of Pennsylvania, Pennsylvania 19104}
\date{\today}

\begin{document}
\maketitle

\begin{abstract}
The Hamiltonian describing a system of strongly correlated electrons coupled to dispersionless phonons was solved numerically for a ring of 8 atoms using the density matrix renormalization group (DMRG) method. It was found that electron correlation and electron-phonon coupling compete against each other, and strong electron correlations suppress the charge-ordered insulating state.  This allows extended polarons to form in the strong electron-phonon coupling regime. It is shown that in this regime two polarons may form a pair via phonon mediated charge-fluctuation interaction. Based on the results we propose a novel mechanism of hole pairing. This mechanism could be relevant to High-$T_{c}$ cuprates.
\end{abstract}

\section{Introduction}
The role of lattice in the mechanism of the High-$T_{c}$ superconductivity in cuprates remains controversial. Even though the majority of researchers believe in magnetic mechanisms because of strong electron correlations and the lack of evidence of a strong isotope effect \cite{Ander87,Emery97}, the growing volume of experimental data \cite{EgB96,Petrov99.2} demonstrating the existence of many anomalous phonon features in cuprates support strong phonon involvement. It is possible that hole pairs in cuprates are formed through a \emph{cooperative} action of strong electron correlations \emph{and} electron-phonon coupling. The purpose of this work is to demonstrate such cooperative effects by numerically studying a model system of strongly correlated electrons coupled to dynamic phonons.

In the simplest form such a system is described by the one-band Holstein-Hubbard Hamiltonian:
\begin{equation}
\label{1bHH}
H_{H-H}=-t\sum_{<ij>\sigma}(c_{i\sigma}^\dagger c_{j\sigma}+H.c.) + U\sum_i n_{i\uparrow}n_{i\downarrow} - g\sum_i (b_i^\dagger + b_i)n_i + \hbar\omega \sum_i (b_i^\dagger b_i + \frac{1}{2}), 
\end{equation}
where $c_{i\sigma}^\dagger$ and $b_i^\dagger$ stand for electron with spin $\sigma$ and phonon creation operators correspondingly, and $n_i=n_{i\uparrow}+n_{i\downarrow}$. 

Since except for a two-site problem \cite{Ran92} no exact solution is known for this model, it has been studied mainly by numerical methods, but the results obtained so far are not directly applicable to High-$T_{c}$ cuprates. The one-polaron problem solved either by exact diagonalization (ED) on small clusters \cite{Alex94,St97}, Monte Carlo method \cite{DeR82,DeR83}, dynamical mean-field theory \cite{Ciuchi97}, or by a special resummation of the strong-coupling perturbation theory \cite{St96} lacks correlation effects. The solution of the two-polaron problem obtained by ED for 1D and 2D clusters \cite{Fehs96} also is not relevant to High-$T_{c}$ cuprates, since it accounts for correlation effects between just two electrons. To describe these materials realistically one has to include the spin background, i.e. carry out calculations for the system near half-filling. Up to now such study was carried out only in the infinite dimensional limit (dynamical mean-field approximation) \cite{Freer93,Freer95,Millis96}. It is not clear if these results can be applied to high-$T_c$ cuprates, which are essentially low-dimensional systems.
 
Since the amount of calculations grows exponentially with the number of interacting particles, so far this case was studied only in approximation of adiabatic ($\hbar\omega \ll t$) and antiadiabatic ($\hbar\omega \gg t$) phonons. In the latter regime $H_{H-H}$ can be transformed by the Lang-Firsov transformation and reformulated in terms of small polarons \cite{Ran93}, but the results are not applicable to High-$T_{c}$ cuprates for which $t \sim 1$ eV. An interesting modification of the Lang-Firsov transformation was employed by Fehske \textit{et al.} \cite{Fehs94.1,Fehs94.2,Fehs95} to study the aspects of both adiabatic and antiadiabatic regimes simultaneously.

In the adiabatic regime phonons are treated as ``frozen'' distortions of the lattice, which can be relaxed into the lowest energy state by iterative diagonalization of the electronic part of the Hamiltonian. Such systems studied by ED include \textit{t-J} model ($\sqrt{10}\times\sqrt{10}$ cluster \cite{Fehs94.3}), a two-band version of \textit{t-J} model ($Cu_{16}O_{32}$ cluster \cite{Yury98}), the three-band Hubbard model ($Cu_{4}O_8$ cluster \cite{Loren94}) and the one-band Hubbard model ($\sqrt{10}\times\sqrt{10}$ cluster \cite{Zhong92}). Also few mean-field calculations on larger clusters \cite{Zaa94,Bish93} were carried out. It is clear, though, that the adiabatic approximation misses the dynamical aspect of electron-phonon interaction in High-$T_{c}$ cuprates. 

The goal of the present work was to study generic effects of electron-phonon coupling in a strongly correlated system near half-filling. To this end a small one-dimensional cluster containig electrons interacting with dynamical phonons has been studied numerically. We found that a new mechanism of hole pairing based on charge fluctuations mediated by phonons arises in this regime. We beleive that it could be a driving force behind the superconductivity in cuprates, although a proper study of size and dimensionality effects is required to establish the magnitude of the effect. This is left for a further study. 

\section{Method}
The Density Matrix Renormalization Group (DMRG) method~\cite{White93} was applied to a 8-site ring cluster at half-filling, with one and two doped holes, and up to 5 phonons per site. The electron-phonon coupling term in~(\ref{1bHH}) was somewhat modified as
\begin{equation}
\label{HHc}
-g\sum_i (b_i^\dagger + b_i) \tilde{n}_i,
\end{equation}
where $\tilde{n}_i=(1-n_{i\uparrow})(1-n_{i\downarrow})$, so that the term describes coupling of phonons only to a \emph{vacant} site due to a doped hole or charge-transfer fluctuations of the spin background. Although this definition neglects the difference in the site being occupied by one or two electrons, it retains the essential feature of the Fr\"{o}lich term in~(\ref{1bHH}): coupling of phonons to background charge fluctuations. This satisfies the goal of the study, and at the same time results in a sizable decrease of required computations due to lower phonon density. 

The system was studied for a broad range of $U$ (4 - 16 eV) and $g$ (0 - 6 eV) parameters, with $t=1$ eV and $\hbar\omega$ fixed at 0.1 eV for most of the calculations. This phonon energy was chosen as a characteristic for LO phonons in High-$T_{c}$ cuprates that cause charge transfer from one copper ion to the neighboring copper ion, and show anomalous dependence on the momentum and temperature \cite{Mcq99,Petr00}. Since in the real system the total number of phonons is not limited, the maximum number $N^{ph}_{max}$ of 5 phonons per site used in this work is clearly a serious limitation. Unfortunately, the increase of $N^{ph}_{max}$ results in a rapid growth of the consumed computer memory. The effect of the phonon space truncation has been monitored by measuring the average number of phonons per site. For large $g/\hbar\omega$ values it reaches a saturation value, which is $\sim 3.3$ for $N^{ph}_{max} = 5$ and therefore results for this region are only of a partial merit. On the other hand, the most important results (hole pairing regime) was obtained for the small-$g$ sector of the parameter space, where effects of phonon truncation are not severe and allow qualitative conclusions be drawn for that part.     

\section{Results and Discussion}
\subsection{Competition between electron correlation and electron-phonon interaction}
\begin{figure}[tb]
   \centering 
   \epsfxsize=8cm
   \epsffile{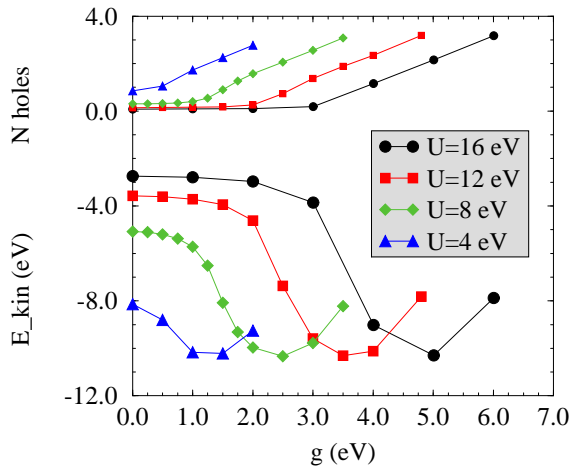}
   \caption{Kinetic energy and the total number of holes (small symbols) for the undoped system.}
   \label{f:kine}
\end{figure}

The kinetic energy ($t$-term in~(\ref{1bHH})) and the total number of holes in the undoped system are shown in Fig.~\ref{f:kine} for different values of $U$ and $g$. 
A rapid decrease in the kinetic energy and a simultaneous increase in the density of holes, defined as a site with no electrons ($\tilde{n}_i=1$), as the value of g is increased suggests that a crossover of the upper Hubbard band (UHB) and the lower Hubbard band (LHB) is taking place. As the two bands overlap, the charge transfer excitations create doubly occupied sites and the number of holes starts to grow rapidly as illustrated by small symbols in Fig.~\ref{f:kine}. For each $U$ the overlap $g$ value can be loosely set to the value at which the distinctive bend in the total number of holes in Fig.~\ref{f:kine} occurs.

\begin{figure}[tb]
   \centering 
   \epsfxsize=8cm
   \epsffile{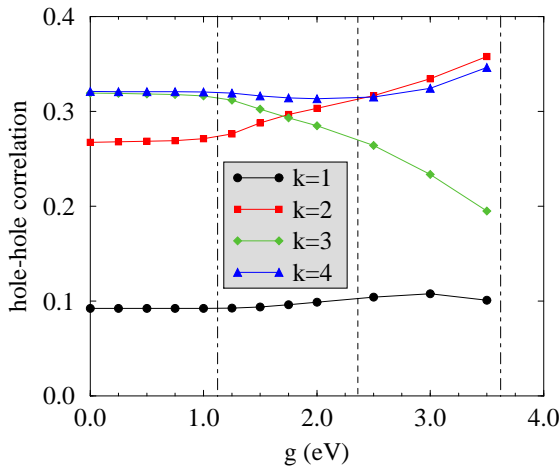}
   \caption{Hole-hole correlation function C(k) for $U=8$ eV. $k=1$ stands for two holes on neighboring sites, 2 for two holes on the next-neighboring sites, etc. Since the total number of holes grows with $g$, the $\sum_k C(k)$ is normalized to 1 for each $g$ value for illustrative purposes. Vertical lines mark the suggested UHB --- LHB overlap region.}
   \label{f:hcorr}
\end{figure}

 After a sharp decrease the kinetic energy rises again, indicating formation of a state with a smaller degree of charge mobility. This happens as UHB sinks below LHB, which corresponds to the charge condensating in the form of bipolaron lattice or charge-density wave (CDW) state. This is confirmed by the hole-hole correlation function defined by 
$$ C(k) = \left\langle \frac{1}{2}\sum_i \tilde{n}_i \tilde{n}_{i+k} \right\rangle $$
shown in Fig.~\ref{f:hcorr} for the undoped system. 
Indeed, as soon as the kinetic energy approaches the minimum, configurations with holes on every other chain site ($k=2$ and $k=4$) gain weight, while the weight of the intermediate hole-hole correlation configuration ($k=3$) decreases, which reveals the rapid onset of the charge ordering. We like to stress here that the appearance of the charge-ordered low-mobility state is delayed up to large values of $g$ in the presence of the on-site correlation energy $U$, and the delay is proportional to $U$. In other words, the electron correlations compete against the hole-electron interaction and suppress polaron condensation. 

Thus the on-site correlation energy $U$ is effectively reduced by the hole-phonon interaction~(\ref{HHc}). Intuitively, the origin of this renormalization is easily understood. As a charge fluctuation producing double occupancy on one site and an induced hole on another increases the energy of the system by $U$, it also couples to the phononic field via~(\ref{HHc}) thus reducing the effective $U$ to $U_{eff}<U$ (for discussion see also \cite{Freer95}). The minima of the kinetic energy curves correspond to the maximum UHB --- LHB overlap, and therefore $U_{eff} = 0$. The relation between the effective Hubbard energy $U_{eff}(U,g)$ and $g$ depends on the $g/\hbar\omega$ ratio and is nonlinear, since the increase in $g$ will cause the increase in the average number of phonons per site and, consequently, the $g$ dependence. However, as can be seen from Fig.~\ref{f:kine}, the value of $U$ corresponding to $U_{eff} = 0$ increases proportionally to $g$, i.e. the relation is approximately linear for this system. We beleive that this is an artifact of the $N^{ph}_{max}$ truncation as discussed above, since it has been found that the phonon density saturates shortly before the kinetic energy minimum, and the regime at the minimum and above does not allow the number of phonons in the system to grow along with $g$. 

\begin{figure}[tb]
   \centering 
   \epsfxsize=8cm
   \epsffile{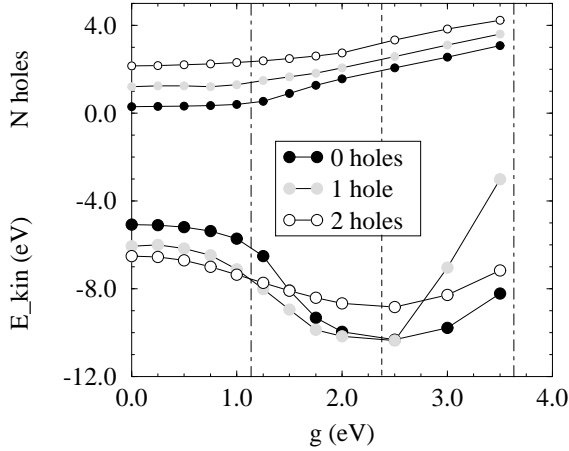}
   \caption{Kinetic energy and the total number of holes for the system with 0, 1 and 2 doped holes ($U=8$ eV). }
   \label{f:kine2}
\end{figure}
\subsection{Nature of the doped hole: Extended polaron formation}
\begin{figure}[tb]
   \centering 
   \epsfxsize=8cm
   \epsffile{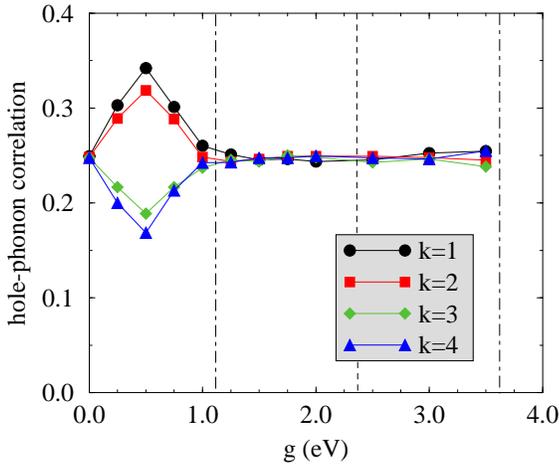}
   \caption{Hole-phonon density correlation function P(k) for $U=8$ eV. The $\sum_k P(k)$ is normalized to 1 for each $g$ value.}
   \label{f:hphcorr}
\end{figure}

We now turn to the system doped with one hole. One can notice in Fig.~\ref{f:kine2}, that the kinetic energy of the one-hole doped system increases slightly as hole-phonon interaction is turned on, suggesting localization tendency of holes, apparently due to phonon drag. The extent of phonon dressing is illustrated by the hole-phonon density correlation function P(k) defined by    
$$ P(k) = \left\langle \frac{1}{2}\sum_i \tilde{n}_i n^{ph}_{i+k} \right\rangle, $$
where $n^{ph}_{i+k}$ stands for the number of phonons at the site $i+k$. P(k) for the system with one doped hole is shown in Fig.~\ref{f:hphcorr}. One can see that for small $g$ values phonons concentrate around the doped hole and the phonon cloud shrinks rapidly as $g$ increases. Since the kinetic energy increases at the same time, it is most likely that a small polaron is forming. As the value of $g$ is further increased, approximately half-way to the UHB --- LHB overlap, a phonon cloud surrounding the doped hole starts to spread again, so that by the time the system reaches the overlap, the cloud is extended over the whole cluster. This is accompanied by a decrease in the kinetic energy, so that this result suggests that an extended polaron is formed as a precursor for the transition into a high-mobility phase at the UHB --- LHB crossover.  

This behavior is very different from that of a conventional Holstein model. It is well known that for the conventional model strong electron-phonon coupling ($g/\hbar\omega > 1$) results in a polaron band collapse (self-trapped polaron) for a single electron, and the CDW formation at half-filling. Here, on the other hand, strong electron correlation competes against the electron-phonon coupling, allowing a phase in which charge becomes mobile before the polaron ordering happens - possibly the extended polaron phase. It is interesting that a similar effect of electron correlations enhancing the conductivity and causing an IM transition was recently found for a disordered Hubbard model in two dimensions~\cite{Dent99}.

\subsection{Charge-fluctuation mechanism of polaron pairing}
\begin{figure}[tb]
   \centering
   \epsfxsize=8cm
   \epsffile{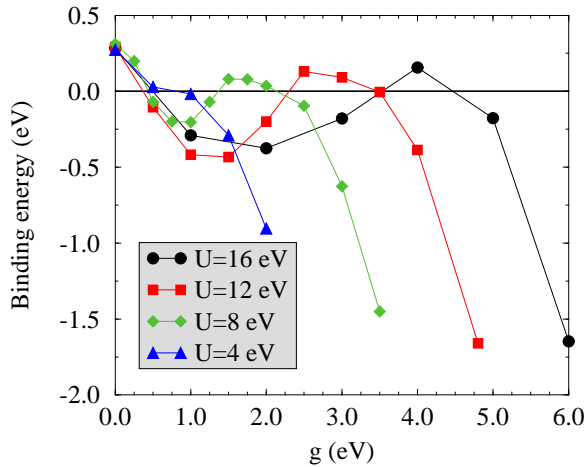}
   \caption{Hole pair binding energy $E_b$.}
   \label{f:binde}
\end{figure}

We now discuss the system with two doped holes. It will be shown that in the range of parameters that corresponds to the extended polaron phase, the two doped holes bind into a pair largerly due to reduction of the background charge fluctuations existing in the phonon cloud surrounding the pair. By defining the hole-pair binding energy as
$$ E_b = E_g(2)+E_g(0)-2E_g(1),$$
where $E_g(n)$ is the ground-state energy for the system doped with $n$ holes, one can study the hole-pair formation and estimate the pairing energy. The obtained binding energy is shown in Fig.~\ref{f:binde} as a function of $g$ and $U$. One can see that the \{$U$, $g$\} parameter space can be divided into four regions. For small values of $g$ $E_b$ is positive, i.e. no hole-pairing occurs. As $g$ increases, $E_b$ becomes negative if $U$ is large enough for the gap between UHB and LHB to exist, and a hole-pair is formed. As $g$ increases futher to the point where the UHB and LHB overlap, the binding energy becomes positive again and the pair formation becomes unfavorable. Finally, as the system crosses over into the charge-ordered region, $E_b$ becomes negative again. Therefore, there are \emph{two} regions of hole-hole attraction --- one in the charge-ordered state, and another in the vicinity of the UHB --- LHB overlap. The first one obviously results in aggregation of doped and induced holes into the charge superlattice. The second one corresponds to the extended polaron regime where the Hubbard gap just closes and the polarons become unstable to pair formation. Clearly, as applied to cuprates the region of interest is the second one, as it has a possibility of resulting in superconductivity.   
\begin{figure}[tb]
   \centering
   \epsfxsize=8cm
   \epsffile{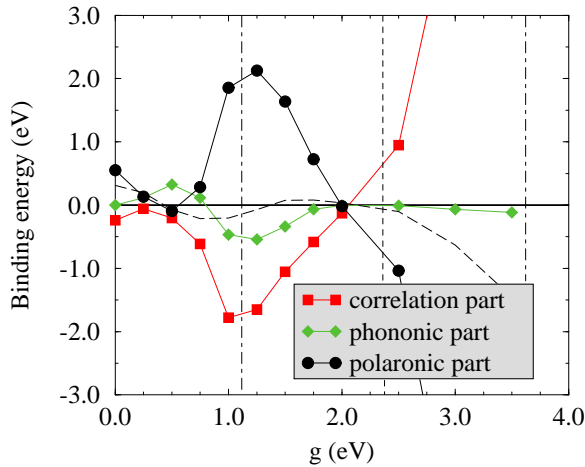}
   \caption{Components of the hole pair binding energy for $U=8$ eV. The binding energy is shown by the dashed curve.}
   \label{f:bind_parts}
\end{figure}
To understand the origin of the binding force in this region we separate $E_b$ into ``polaronic'' ($t$ and $g$ terms), ``phononic'' ($\hbar\omega$ term), and ``correlation'' ($U$ term) inputs, by calculating expectation value of the corresponding terms in the studied Hamiltonian for the ground state. The result for $U=8$ is shown in Fig.~\ref{f:bind_parts}. Two important conclusions follow. First of all, one can see that the correlation energy plays a crucial role in the pair formation mechanism, giving most of the ``binding'' (negative) input into $E_b$. On the contrary, the effect of dynamic ``polaronic'' interaction is to the most part negative or ``anti-binding''. A peculiar dip in the ``polaronic'' curve and corresponding bumps in ``phononic'' and ``correlation'' curves at $g \sim 0.5$ eV correspond to the small polaron --- extended polaron transition mentioned above. In fact, for larger $U$ values the dip in the polaronic binding energy creates a dominant input into the hole-pair binding for small values of $g$, i.e. before the exteded polaron is formed, which corresponds to a conventional small polaron binding mechanism.

Secondly, the behaviour of $E_b$ components clearly correlates with the UHB --- LHB crossover, when the extended polaron becomes unstable. As $g$ increased through the crossover, the polaronic component of $E_b$ remains positive and increases, while the correlation component stays negative and decreases. As the two bands begin to overlap, the polaronic and the correlation parts reach their maximum and minimum correspondingly. When the overlap is at its largest ($U_{eff} \simeq 0$), the two components change signs and diverge as soon as the system reaches the charge-ordered regime. 

From the definition of $E_b$ we see that it compares the energy of two separate doped holes and their backgrounds with the energy of two holes confined into one background plus the undoped background itself. Therefore, the positive sign of the $E_b$ polaronic component, for example, signifies that the system with two separate doped holes allows more hole motion than the system with the two holes bound together. By analyzing Fig.~\ref{f:bind_parts} along these lines we propose the \emph{phonon mediated charge-fluctuation mechanism} of hole pairing as described below. For illustration purposes the model is generalized to two dimensions.

\begin{figure}[tb]
   \centering
   \epsfxsize=8cm
   \epsffile{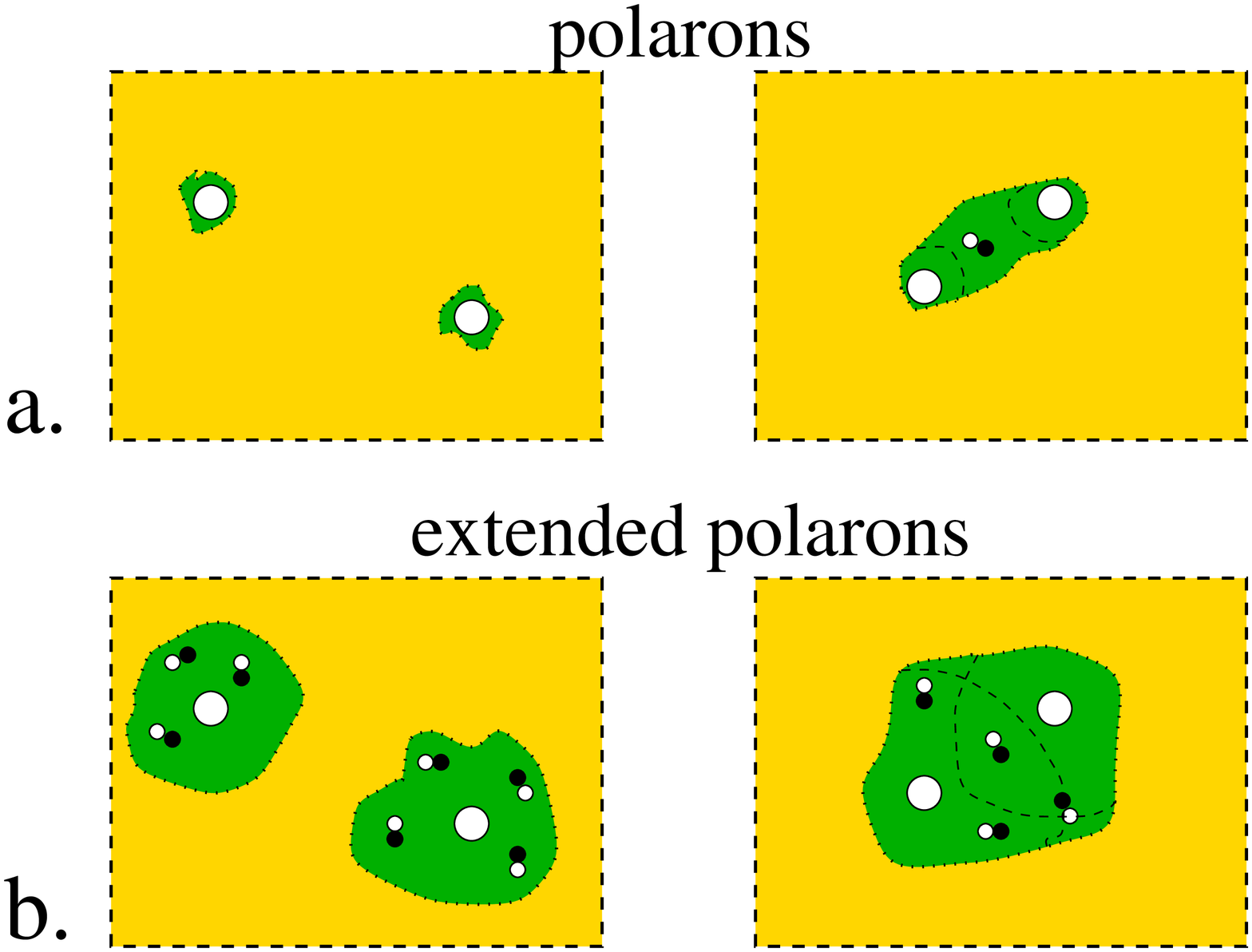}
   \caption{Two regimes of doped holes binding: (a) small polarons, (b) extended polarons, (left) unpaired state, (right) paired state. Large open circles stand for doped holes, small black and white circles for background charge-transfer fluctuations, while shaded regions represent phonon clouds.}
   \label{f:2regimes}
\end{figure}
We start from the state with $g=0$ and $U$ large enough for the Hubbard gap to be open. Since two doped holes confined to one region strongly interfere with each other's motion $E_b$ is positive. As the hole-phonon interaction is turned on, conventional small polarons are formed, and motion of a hole is reduced due to renormalization of its mass. In this regime two holes can \emph{combine} their phononic clouds, thus increasing their mobility as compared to the separated holes (Fig.~\ref{f:2regimes}a). This is a regular polaron coupling mechanism, and it applies to the hole-pairing regime for large $U$ and small $g$ mentioned above. As the $g/\hbar\omega$ ratio is increased the phonon cloud of a single polaron spreads on neighboring sites forming an extended polaron (Fig.~\ref{f:2regimes}b). The role of correlation interaction $U$ becomes crucial here since it delays the charge-ordered state formation and allows extended polarons to form. In this regime holes tend to \emph{share} their phonon clouds, which will decrease their kinetic energy relative to the unpaired holes. Altogether this explains the dip observed in the polaronic component of $E_b$. 

The main reason for extended polarons to share their phonon clouds is the charge-transfer fluctuations of the background electrons. Extended phononic field causes charge-transfer background fluctuations ($\uparrow$, $\downarrow$) $\leftrightarrow$ (0, $\uparrow\downarrow$) via the hole-phonon interaction (see Fig.~\ref{f:2regimes}b). The double ocupancy will cost on-site correlation energy $U$ (in addition to the $\hbar\omega$ energy of the phononic field), and unless the UHB and LHB overlap, the two polarons prefer to share their domains to minimize the number of the (0, $\uparrow\downarrow$) pairs. This is corroborated by large negative values of the correlation ($U$) component of $E_b$ in Fig.~\ref{f:bind_parts}. On the other hand, as the bands overlap, the ``fluctuation'' pair formation costs less and less energy, so that the tendency for polarons to stay together will decrease, until the polaron-pair formation will become unfavorable ($E_b$ becomes positive again in Fig.~\ref{f:binde}). This phonon-induced correlation mechanism of pairing is closely related to the idea of "correlation bag" mechanism proposed by Goodenough \cite{Good90}. 

\begin{figure}[tb]
   \centering
   \epsfxsize=8cm
   \epsffile{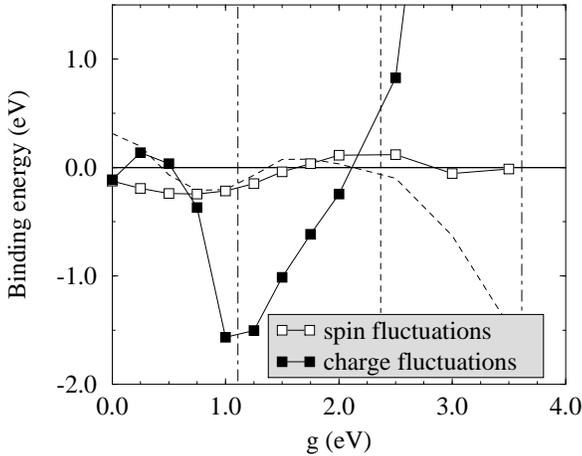}
   \caption{Ising and charge-transfer components of the correlation component of $E_b$ for $U=8$ eV. The binding energy is shown by the dashed curve.}
   \label{f:bind_parts1}
\end{figure}
Since the Hubbard ``correlation'' component $U$ includes both spin-fluctuation (as in the $t-J$ model) and charge-fluctuation parts, it is important to estimate their effect on pair formation separately. To this end we evaluated the ``$t-J$'' part by calulating the ground state expectation value of the Ising term: $\frac{4t^2}{U}\langle \sum_i S^z_i S^z_{i+1} \rangle$. The charge part was defined as the difference between the U term in Fig. 6 and the spin fluctuation contribution. The resulting components are shown in Fig.~\ref{f:bind_parts1}. It can be seen that the spin contribution is dominant in the small-polaron regime as expected, but it becomes weaker as the UHB -- LHB crossover occurs. Instead, the charge-fluctuation input into $E_b$, which is suppressed in the small-polaron regime, quickly becomes dominant as the Hubbard gap decreases and the charge-transfer process is activated. 

\subsection{Relevance to the High-$T_c$ cuprates}
\begin{figure}[tb]
   \centering
   \epsfxsize=8cm
   \epsffile{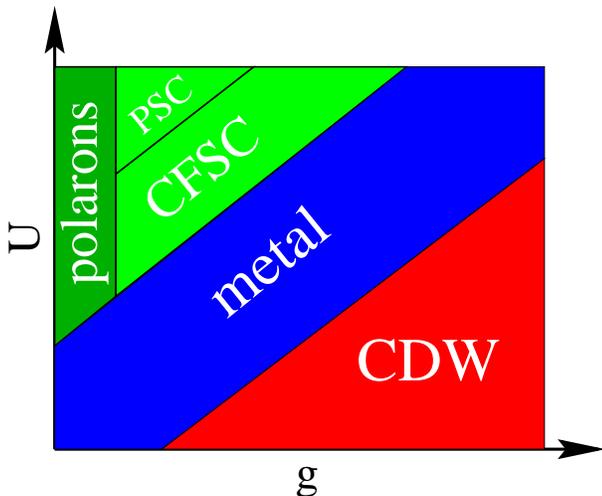}
   \caption{A generic U-g phase diagram for the one-band Hubbard-Holstein model. PSC stands for polaronic superconductivity, and CFSC for phonon mediated charge-fluctuation superconductivity.}
   \label{f:U-g}
\end{figure}
A schematic U-g phase diagram in Fig.~\ref{f:U-g} summarizes the qualitative picture obtained in this work in the form of suggested behavior of a real physical system (a high-$T_c$ cuprate, for example). The relative position of the UHB and the LHB determines its electronic behavior. When the Hubbard gap is open (large $U$, small $g$), the system is a doped insulator with polaronic conduction. When $U$ is small, or the effects of $U$ and $g$ are competing, the UHB and LHB overlap and the system is metallic. For small $U$ and large $g$ the insulating bipolaronic-lattice/CDW regime prevails. The superconducting transition happens as polarons become unstable to pair formation. For large $U$ values the pairing mechanism is the conventional polaronic coupling, while for lower $U$ values in the vicinity of the polaron-metal crossover the mechanism is that of phonon mediated charge-fluctuations described above. While the superconducting transition temperature in the polaronic regime is expected to be low due to the strong mass renormalization, in the polaron-metal crossover regime the spread of a polaron would result in a much smaller mass, and thus could result in a high transition temperature.

Qualitatively it is clear that the proposed charge-fluctuation mechanism will produce the maximum pair-binding energy at the Insulator-to-Metal (IM) crossover, as seen in Fig.~\ref{f:bind_parts}. Indeed, by suppressing the superconductivity by magnetic field it has been shown that the IM crossover happens at the same doping as the maximum $T_c$ is reached ~\cite{Boeb96}.

The doping driven IM transition in LSCO suggests a percolative origin of conductivity at low temperatures. In fact, such a mechanism involving percolation between large polarons existing at low temperatures was proposed to explain IM transition and CMR phenomenon in manganites~\cite{EgamiM96,Louca99}. The present results demonstrate that the large (extended) polaron regime can arise naturally in doped Mott-Hubbard insulators with strong electron-phonon coupling. As the number of doped holes increases extended polarons overlap, and the percolative conductance developes. Furthermore it was observed for manganites that local dynamic coupling of hole and phonon (vibronic state) develops at the IM transition~\cite{Louca99}. In cuprates such local dynamic coupling could lead to superconductivity of extended polarons. The anomalous behavior of LO phonons recently observed by neutron inelastic scattering could be the manifestation of such coupling \cite{Mcq99,Petr00}.

\section{Conclusion}
By solving the one-band Hubbard hamiltonian coupled to dynamic phonons by a Fr\"{o}lich-type term near half-filling several novel observations have been made: 1) Electron correlation and electron-phonon coupling compete with each other. Thus the charge-ordered insulating state is suppressed by electron correlations, and extended polaron formation becomes possible for strong electron-phonon coupling. 2) The extended polarons formed by doped holes bind into pairs, and their binding energy is strongly enhanced by electron correlation contribution. Based upon these observations we propose that phonon mediated pairing interaction based upon charge fluctuations in the half-filled system can be a relevant mechanism of superconductivity in High-$T_{c}$ cuprates.  

\section*{Acknowledgments}
The authors are grateful to A. R. Bishop, H. R\"{o}der, T. Gammel and V. Emery for useful discussions. The work was supported by the National Science Foundation through DMR96-28134.

%\bibliographystyle{/home/petrov/DOCS/REVTEX/prsty}
%\bibliography{/home/petrov/DOCS/THESIS/base}

\begin{thebibliography}{10}

\bibitem{Ander87}
P.~W. Anderson, Science {\bf 235},  1196  (1987).

\bibitem{Emery97}
V.~J. Emery, S.~A. Kivelson, and O. Zahar, Phys. Rev. B {\bf 56},  6120
  (1997).

\bibitem{EgB96}
T. Egami and S.~J.~L. Bilinge,  in {\em Physical Properties of High-Temperature
  Superconductors}, edited by D. Ginsberg (World Scientific, Singapore, 1996),
  Vol.~V, pp.\ 265--373.

\bibitem{Petrov99.2}
T. Egami {\it et~al.},  in {\em High temperature superconductivity}, edited by
  S.~E. Barnes and et. al. (American Institute of Physics, Woodbury, New York,
  1999), pp.\ 231--236.

\bibitem{Ran92}
J. Ranninger and U. Thibblin, Phys. Rev. B {\bf 45},  7730  (1992).

\bibitem{Alex94}
A.~S. Alexandrov, V.~V. Kabanov, and D.~K. Ray, Phys. Rev. B {\bf 49},  9915
  (1994).

\bibitem{St97}
M. Capone, W. Stephan, and M. Grilli, Phys. Rev. B {\bf 56},  4484  (1997).

\bibitem{DeR82}
H. De~Raedt and A. Lagendijk, Phys. Rev. Lett. {\bf 49},  1522  (1982).

\bibitem{DeR83}
H. De~Raedt and A. Lagendijk, Phys. Rev. B {\bf 27},  6097  (1983).

\bibitem{Ciuchi97}
S. Ciuchi, F. Pasquale, S. Fratini, and D. Feinberg, Phys. Rev. B {\bf 56},
  4494   (1997).

\bibitem{St96}
W. Stephan, Phys. Rev. B {\bf 54},  8981  (1996).

\bibitem{Fehs96}
G. Wellein, H. R\"{o}der, and H. Fehske, Phys. Rev. B {\bf 53},  9666  (1996).

\bibitem{Freer93}
J.~K. Freericks, M. Jarrell, and D.~J. Scalapino, Phys. Rev. B {\bf 48},  6302
  (1993).

\bibitem{Freer95}
J.~K. Freericks and M. Jarrell, Phys. Rev. Lett. {\bf 75},  2570   (1995).

\bibitem{Millis96}
A.~J. Millis, R. Mueller, and B.~I. Shraiman, Phys. Rev. B {\bf 54},  5389
  (1996).

\bibitem{Ran93}
J. Ranninger, Phys. Rev. B {\bf 48},  13166  (1993).

\bibitem{Fehs94.1}
U. Trapper, M. Fehske, H.and~Deeg, and H. B\"{u}ttner, Z. Phys. B {\bf 93},
  465  (1994).

\bibitem{Fehs94.2}
H. Fehske {\it et~al.}, Z. Phys. B {\bf 94},  91  (1994).

\bibitem{Fehs95}
F. H., H. R\"{o}der, and G. Wellein, Phys. Rev. B {\bf 51},  16582  (1995).

\bibitem{Fehs94.3}
H. R\"{o}der, H. Fehske, and R.~N. Siver, Europhys. Lett. {\bf 28},  257
  (1994).

\bibitem{Yury98}
Y. Petrov and T. Egami, Phys. Rev. B {\bf 58},  9485  (1998).

\bibitem{Loren94}
J. Lorenzana and A. Dobry, Phys. Rev. B {\bf 50},  16094  (1994).

\bibitem{Zhong92}
J. Zhong and H.-B. Schl\"{u}tter, Phys. Rev. Lett. {\bf 69},  1600  (1992).

\bibitem{Zaa94}
J. Zaanen and P.~B. Littlewood, Phys. Rev. B {\bf 50},  7222  (1994).

\bibitem{Bish93}
K. Yonemitsu and A.~R. Bishop, Phys. Rev. B {\bf 47},  8065  (1993).

\bibitem{White93}
S.~R. White, Phys. Rev. B {\bf 48},  10345  (1993).

\bibitem{Mcq99}
R. J. McQueeney, Y. Petrov, T. Egami, M. Yethiraj, G. Shirane and Y. Endoh, Phys. Rev. Lett. {\bf 82}, 628 (1999). 

\bibitem{Petr00}
Y. Petrov, T. Egami, R. J. McQueeney, M. Yethiraj, H. A. Mook and F. Dogan, cond-matt/0003414.

\bibitem{Dent99}
P.~J.~H. Denteneer, R.~T. Scalettar, and N. Trivedi, Phys. Rev. Lett. {\bf 83},
   4610  (1999).

\bibitem{Boeb96}
G.~S. Boebinger, Y. Ando, and A. Passner, Phys. Rev. Lett. {\bf 77},  5417
  (1996).

\bibitem{EgamiM96}
T. Egami, J. Low Temp. Phys. {\bf 105},  791  (1996).

\bibitem{Louca99}
D. Louca and T. Egami, Phys. Rev. B {\bf 59},  6193  (1999).

\bibitem{Good90}
J. B. Goodenough and J. Zhou, Phys. Rev. B {\bf 42}, 4276 (1990).


\end{thebibliography}

\end{document}